# The fiber-fed preslit of GIANO at T.N.G.

A. Tozzi[*a], E. Oliva[a], L. Origlia[b], C. Baffa[a], V. Biliotti[a], G. Falcini[a],
E. Giani[a], M. Iuzzolino[a], F. Massi[a], N. Sanna[a], S. Scuderi[c], M. Sozzi[a].
[a]INAF – Osservatorio Astrofisico di Arcetri, L.go E. Fermi 5, 50125 Firenze, Italy (E.U.);
[b]INAF – Osservatorio Astronomico di Bologna, Via Ranzani 1, 40127 Bologna, Italy (E.U.);
[c]INAF – Osservatorio Astronomico di Catania, Via S.Sofia 78, 95123 Catania, Italy (E.U.).

## ABSTRACT

Giano is a Cryogenic Spectrograph located in T.N.G. (Spain) and commisioned in 2013. It works in the range 950-2500 nm with a resolving power of 50000.
This instrument was designed and built for direct feeding from the telescope [2]. However, due to constraints imposed on the telescope interfacing during the pre-commissioning phase, it had to be positioned on the rotating building, far from the telescope focus. Therefore, a new interface to the telescope, based on IR-transmitting ZBLAN fibers with 85μm core, was developed.Originally designed to work directly at the f/11 nasmyth focus of the telescope, in 2011 it has decided to use a fiber to feed it.
The beam from the telescope is focused on a double fiber boundle by a Preslit Optical Bench attached to the Nasmith A interface of the telescope. This Optical Bench contains the fiber feeding system and other important features as a guiding system, a fiber viewer, a fiber feed calibration lamp and a nodding facility between the two fibers. The use of two fibers allow us to have in the echellogram two spectrograms side by side in the same acquisition: one of the star and the other of the sky or simultaneously to have the star and a calibration lamp. Before entering the cryostat the light from the fiber is collectd by a second Preslit Optical Bench attached directly to the Giano cryostat: on this bench the correct f-number to illuminate the cold stop is generated and on the same bench is placed an image slicer to increase the efficiency of the system.

## 1. INTRODUCTION

During the Giano's cryostat commissioning on July 2012 the spectrograph was located in the TNG Nasmyth A and was positioned on the floor of the rotating building, detached from the telescope.
This is the reason why Giano[1][3] is feed via special fibers with extended transmission to the infrared wavelengths which are interfaced to the telescope at the f/11 OIG focus of the tlescope.
The Giano's preslit system is composed by two different optical benches: one mounted on the OIG Telescope interface, and the other mounted directly on one side of the Giano Cryostat, in front of the input windows. They have been realized mainly using commercial components in particular for the optomechanical devices, this because of the short time used for the assembling.

PRESLIT OIG BOX:
1) a full controlled telescope's optical axis - fiber's axis alignment thanks to the available visible pupil check, the adjustable optical components inside the OIG board and the possibility to align the whole OIG box using seven external micrometers realizing a full 3D alignment system.
2) The possibility to look at the fibers position on the guider camera. This is the upstream mode that is referred to the light going from a calibration source to the guider camera through the fibers (in order to check the fiber alignment with respect to the telescope's axis).

PRESLIT GIANO BOX

---

[*] atozzi@arcetri.astro.it; phone +390552752205

1) a full controlled Giano optical bench - fiber's axis alignment thanks to the adjustable optical components inside the Giano side box board and the possibility to align the whole box using seven external micrometers realizing a full 3D alignment system.
2) The possibility to increase the efficiency using a slicer prism
3) The possibility to look at an intermediate focus before the beam enters in the cryostat.

## 2. THE FIBERS-TELESCOPE INTERFACE

The fiber to telescope interface is realized by the Preslit OIG Box, described in the following paragraphs.
First of all it is important to remember that the two Preslit boxes are optically connected by a couple of fibers optics, assembled as shown in the following design, in the same SMA connector.

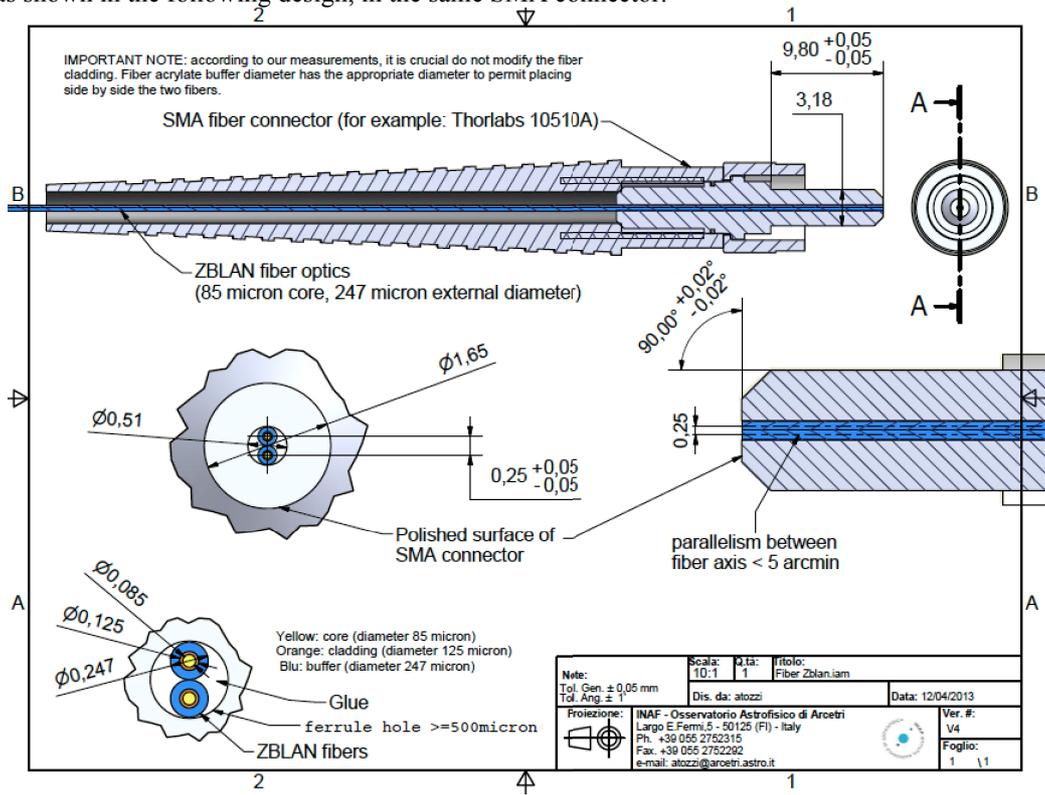

Figure 1 Design of the custom fiber optics patch cord. Core diameter of each fiber is 84 micron (1 arcsec on sky). Distance between centers is 250 micron (3 arcsec on sky).

The two fiber terminations are aligned vertically with respect to the Preslit OIG box optical bench plane: this will be important for the definition of the observing modes as described in chapter 2.6 "Observing modes". The fibelrs are madi in ZBLAN ant their core dimension is enough to accept one arcsec on sky, while the distance between centers is 3 arcsec on sky. This means that the core is 85 micron and their interdistance is 250 micron.
Finally, it is fundamental to underline that the axis of the fiber must be mechanically aligned relative to the optical axis of the pre-slit optics and with respect the telescope optical axis.

### 2.1 Opto-mechanical design

The OIG box is based on a commercial 60X30 cm aluminum optical bench by Thorlabs onto which the optics are positioned as visible in Figure 2 and Figure 3. The opto-mechanical design is based on commercial elements including, in particular, three off-axis parabolic mirrors. The Oig box can be aligned using six micrometers, two of them visible in the following figure. The alignment procedure consists into the preliminary alignment of the internal optical componts of the OIG box, that can be realized in laboratory, and after this into place the box on the mechanical support structure located at the OIG telescope interface. Then the micrometers can be adjusted to align the optical axis of the OIG Box with respect the incoming f/11 optical beam from the telescope. Focus and pupil position can be regulated, too.

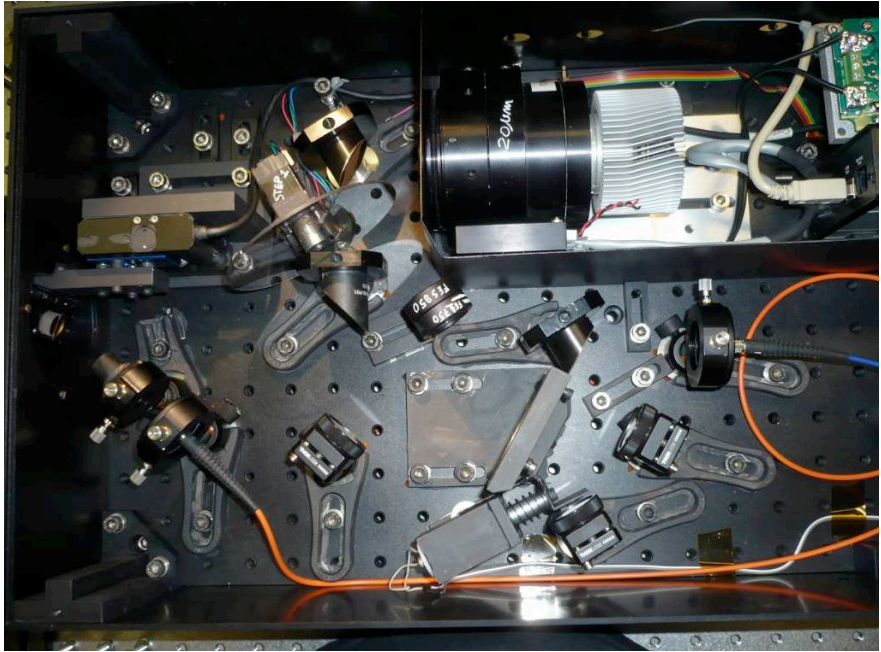

Figure 2 Optics of the Preslit OIG box.

In Figure 4 and Figure 5 are represented the conceptual designs of the Preslit OIG Box. The optical design is composed by at least four optical designs, described in the following paragraphs. This because of the different operating modality we have for this board.

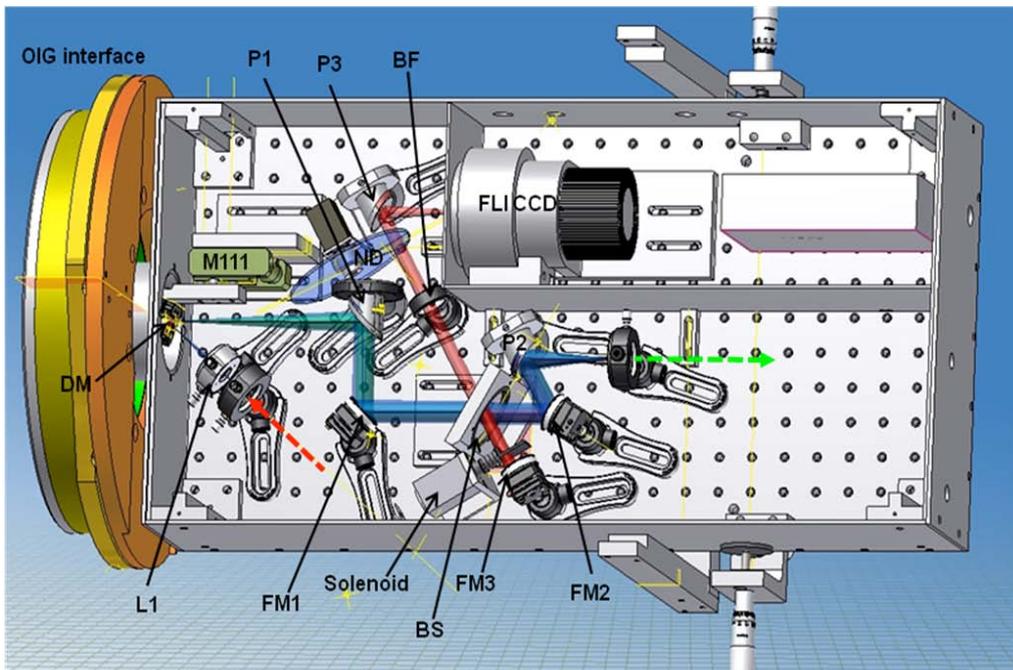

Figure 3 Inventor design of the Preslit OIG box. FM are Folding plane Mirrors, P are parabolic mirrors, BS is the dichroic Beam Splitter, ND the neutral Density sector wheel, DM the D shape Mirror, L1 is a lens, BF a Pass Band filter. In dashed red arrow the input of the fiber calibration light, in green the science output fiber.

In particularly in Figure 4 the observing configuration in which one fiber looks at the sky and the other looks at the calibration lamp is schematically shown., while in Figure 5 the Upstream modality is represented: an Infrared LED is remotely switched on in the Preslit Giano Box and placed in front of the fibers using a motorized system. This is Fiber Viewer modality.

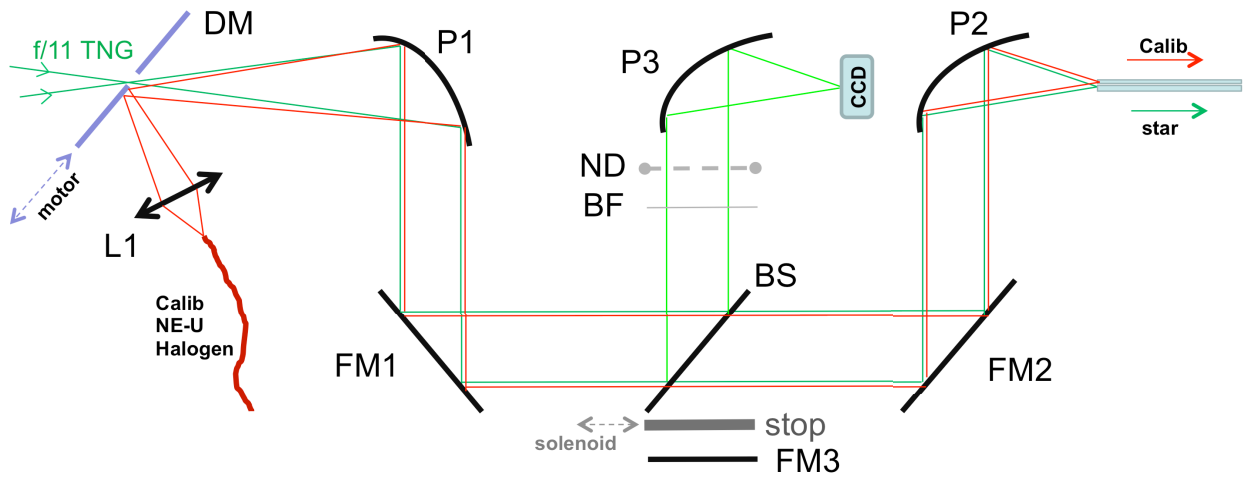

Figure 4 Conceptual plot of the Preslit OIG box in the downstream modality.
Dashed lines represent motorized axis.

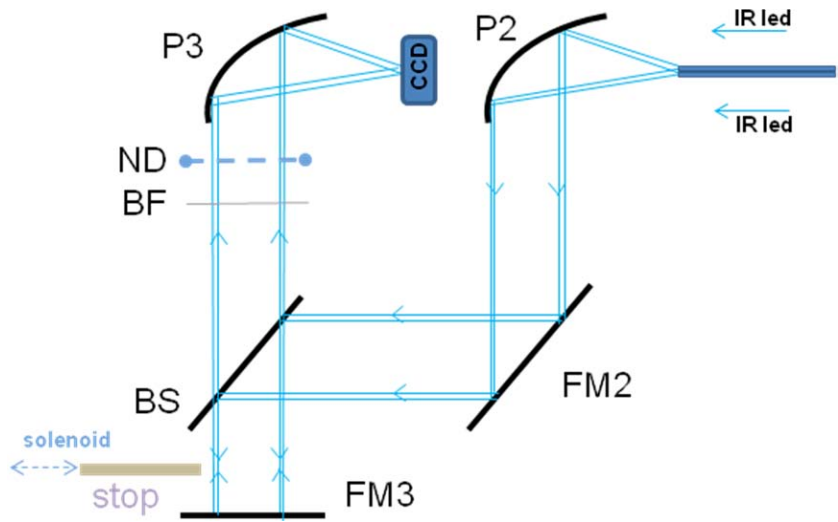

Figure 5 Conceptual plot of the Preslit OIG box in the upstream modality.
P3 is an identical off axis parabola as P2.

In Figure 4 and Figure 5 the dashed lines represent the movable axis, remotely controlled, that are:

1) The motor stage M-111 for the position regulation of the D shape mirrors located on f/11 TNG focus based on a Mercury PI commercial driver,
2) The motor stage for the Neutral Density filter wheel, based on a custom step motor driver,
3) The solenoid to insert/remove the light stop for the fiber viewer.

The Zemax designs for the different optical path are below listed and described.

## 2.2 Light path description in the downstream working mode

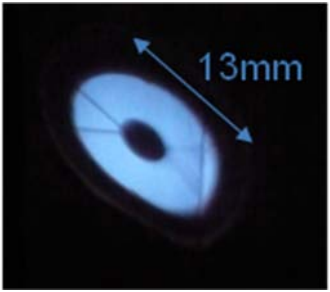

The F/11 beam from the telescope is collimated and re-imaged at the fiber F/4.9 focus using two off-axis parabolae and two bending mirrors. The output F/4.9 beam has a diffraction-limit optical quality and is pseudo telecentric: nominally the exit pupil in is located 151 mm far from the focus and has a diameter of 31 mm, but the chief ray angle for a field of 2 arcsec in sky far from on axis one is only 3,5 arcmin. This implies a lateral shift of the pupil of 0.017%. A dichroic beam-splitter on the collimated beam reflects the blue (<850nm) light to the guider. An image of the TNG pupil is accessible before the dichroic as visible in the photo on this side taken during daytime placing a piece of paper in front of the dichroic. The dome was opened during the test so the primary mirror and spiders were well visible.

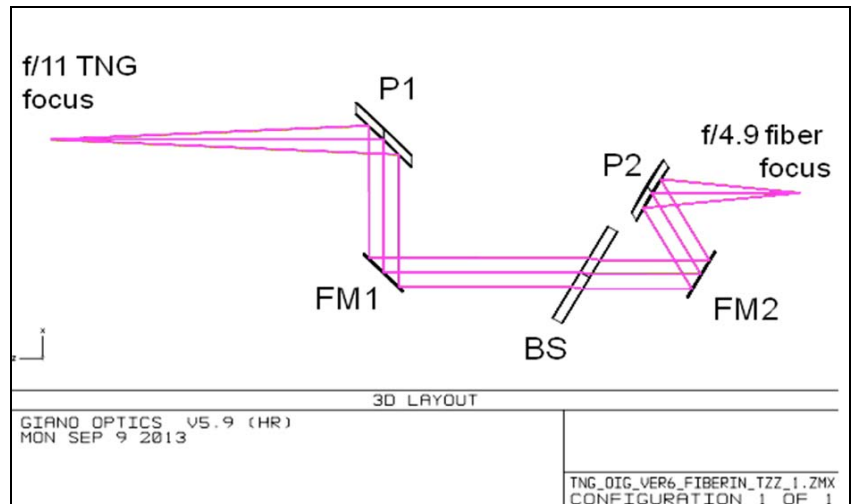

Figure 6 Zemax optical design of the downstream mode. P1 and P2 are commercial parabolas, BS is the dichroic beamsplitter, FM are folding mirrors.

Figure 7 shows the design of the downstream mode. The first off axis parabola, P1,is placed 152.4 mm after the f/11 TNG focus so the output beam is collimated by the P1 parabolic mirror, having an Effective Focal Length (EFL) of 152,40 mm with an error of ±1%. The beam is folded and a pupil image is placed in the dichroic beam splitter. So P1 realizes a real image of the primary mirror on the beam splitter. The second off axis parabola, P2, has an EFL of 67,74 mm (with an error of ±1%) and generates and f/4.9 output beam focalized on the input fiber termination. The scale of this focus is 83,4 micron/arcsec in sky and the whole system is diffraction limited in the full wavelength range 900-2400 nm. The following table shows the scale factor of the different foci present in Preslit OIG Box:

| Focus Position | f/number | Scale [micron/arsec sky] |
|---|---|---|
| TNG focus on OIG flange | f/11 | 187,0 |
| Fiber In on ZBLAN fibers | f/4,9 | 83,4 |
| CCD Guider plane | f/4,9 | 83,4 |

Table 1 Focus scale factor for the Preslit OIG box

The magnification factor from f/11 TNG focus to f/4.9 Fiber focus is -2.24.

## 2.3 Light path description in the calibration mode.

The calibration mode optical design is represented in Figure 7. A fiber optics of 400 micron core diameter in ZBLAN feeds the Preslit OIG box: the calibration NE-U and Halogen lamps are located in a rackmount, placed on the Nasmity A platform. The light is refocused by L1 (commercial CaF singlet of 15 mm focal length) on the f/11 nominal position of the TNG focus where a D shape mirror is located and can be easily positioned using a remote controlled motor linear stage (M111 of MICOS/PI). Then the light follows the same optical path of the natural star and is focused on the fiber input terminations.

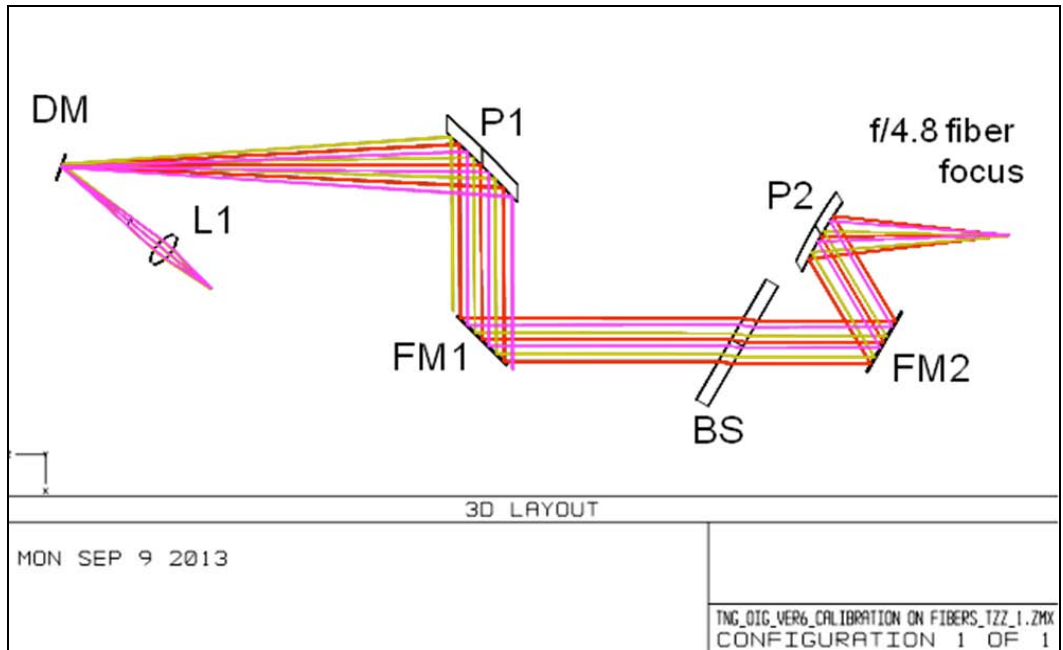

Figure 7 Zemax optical design for the Calibration mode. P1 and P2 are commercial parabolas, BS is the dichroic beamsplitter, FM are folding mirrors, DM is D shape couple of mirrors remotely controlled.

The M111 MICOS/PI translation stage controls the position of two D shape mirrors that are custom assembled to permit different observing operating modes as described in the following chapter.
The final f/number of the calibration beam is approximately f/4 with a scale factor of approximately 1. The nominal magnification from the fiber source (400 micron fiber) to the final focus is equal to one: so the input terminations of the scientific ZBLAN fiber can be fed by the incoming light from the calibration light source, that can be chosen between the Ne-U lamp and the Halogen lamp. The calibration sources are located into the Preslit Rackmount.

## 2.4 Light path description of the guider camera

The Guider is based on a CCD camera made by Finger Lake based on a CCD having 512X512 pixel array of 20 micron pixel size. The focal camera is realized using an off axis parabola (P3) identical to P2. The scale on CCD is 83 micron/arcsec on sky that it means there is a magnification of 0.44 from f/11 TNG focus to the focus on the guider CCD. The nominal FOV of the guider is 2X2 arcmin, but during sky observation the FOV is partially vignetted by the D shape Mirror support located on the f/11 TNG focus and necessary to change the observing mode.

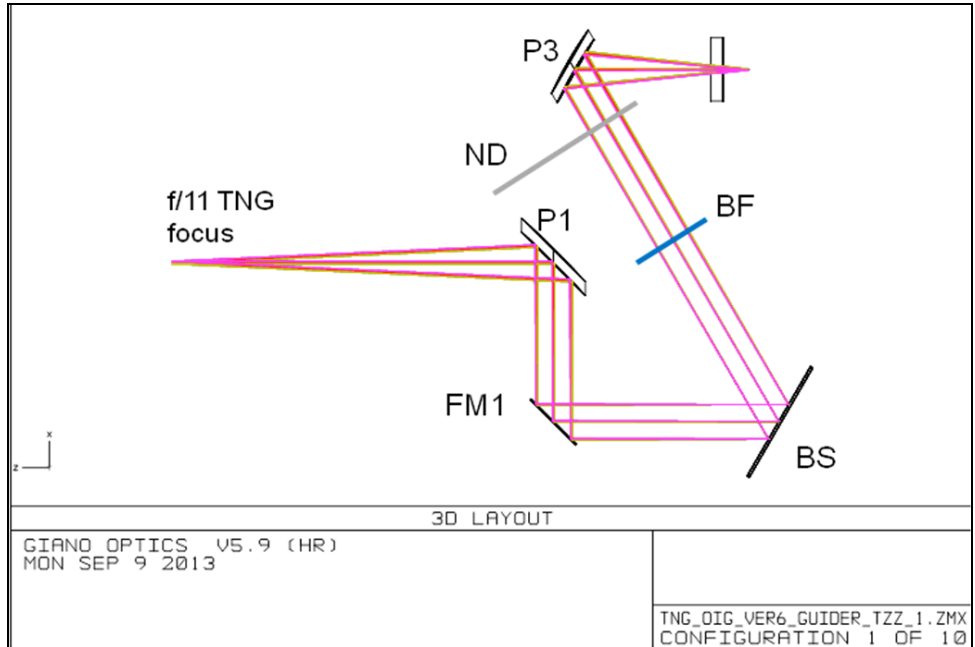

Figure 8 Zemax optical design for the Guider branch. P1 and P3 are commercial parabolas, BS is the dichroic beamsplitter, FM ais a folding mirror. ND is a neutral density filter wheel. BF is a passband filter.
(file: TNG_oig_ver6_guider_tzz_1.zmx)

After the Dichroic Beam Splitter there are positioned a BandPass filter (BF) and a multi position Neutral Density Wheel, having the following ND values: 0.1, 0.2, 0.3, 0.4, 0.5, 0.6, 1.0, 2.0, 3.0 and 4.0. Before the ND filter wheel the Bandpass Filter is placed on the optical path: it has been realized using two commercial filters of thorlabs: cut-on FEL 750, cut-off FES 950.

**2.5 Light path description of the Fiber Viewer camera**

The Fiber Viewer is based on the same CCD camera used by the Guider.
On the optical beam in the Giano Preslit Box an IR LED can be inserted and switched on varying the intensity. The light is partially collected into the two ZBLAN fibers and reaches the Preslit OIG box using the same fibers normally used for the scientific measures on the sky.
The outgoing light from these two fibers is collimated by P2 parabolic mirror on the FM2 folding mirror. Then is partially reflected by the Beam Splitter (BS) and totally reflected back again from FM3 folding mirror. This mirror is normally covered by a black stop placed in front of it, but this cover can be remotely controlled and removed. The light from FM3 passes through the beam splitter BS and follows the same optical path of the incoming light from star, normally reflected by the dichroic beam Splitter BS. On Finger Lake CCD (FLI CCD) an image of the two fibers is generated on the CCD Guider, simultaneously or not with the scientific target.

The image of the two fibers is reimaged by the two identical off axis parabolas P3 and P2 on the CCD guider and so the magnification factor is equal to one. The position of the two fiber images can be recorded and that coordinates on the CCD, F1(x1;y1) and F2(x2;y2), will represent the correct position in which to align the scientific target under investigation during scientific observation.
This is true provided that the folding mirror FM3 has not changed position in the course of time, otherwise a recalibration procedure will be necessary and an offset to the two CCD pixel coordinates F1 and F2 will be necessary.

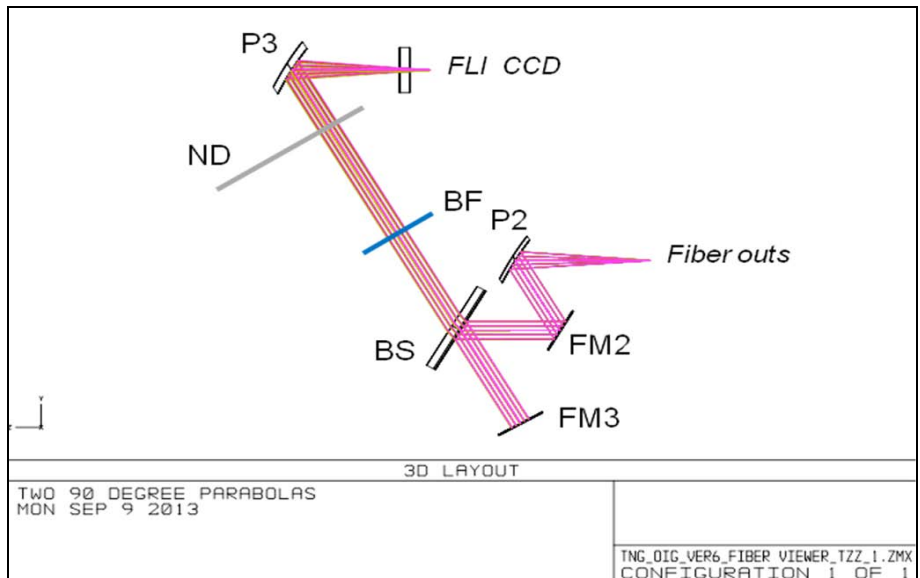

Figure 9 Zemax optical design for the Fiber Viewer branch. P2 and P3 are commercial identical parabolas, BS is the dichroic beamsplitter, FM3 is a folding mirror. ND is a neutral density filter wheel. BF is a passband filter.

### 2.6 Observing modes

We now illustrate the Observing Modes (OM).

The inputs of the two optical ZBLAN fibers are vertically disposed in the Preslit OIG box. This fact permits to operate in different observation modality only moving the translation stage M-111 on which the two D shape mirrors are mounted as visible in the detailed design/photo of Figure 10 and Figure 11.

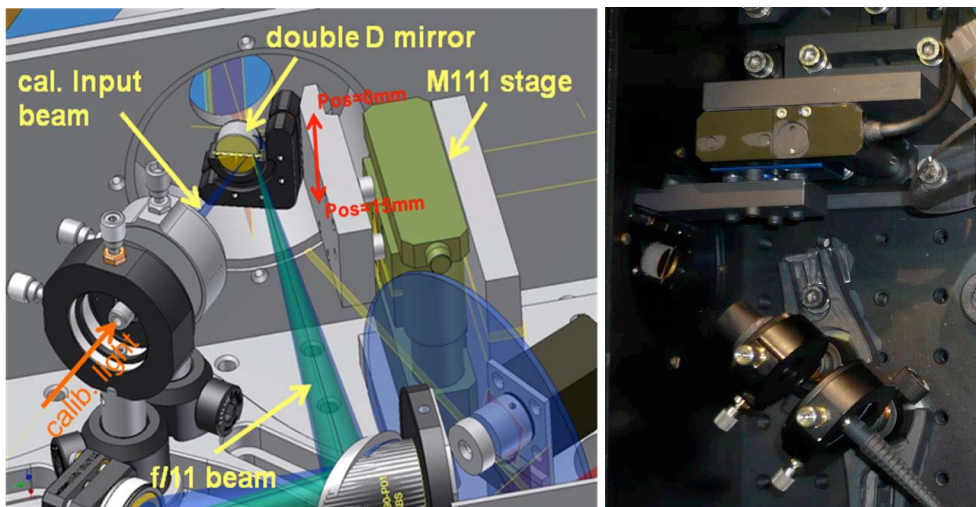

Figure 10 Detailed view of the D mirror, placed on the f/11 TNG focus.

Mirrors are vertically mounted and the shape of these mirrors is like a D where the straight side is very well worked to generate a knife. Two of these mirrors have been mounted on the same mechanical mount using Millbond optical glue: the two parallel sides form a slit of about one millimeter located at the nominal f/11 TNG focus. The tilt regulations of the mechanical mount on which the D mirrors are placed are useful to align the outgoing calibration beam from lens L1, to be overlapped to the f/11 TNG telescope chief ray.

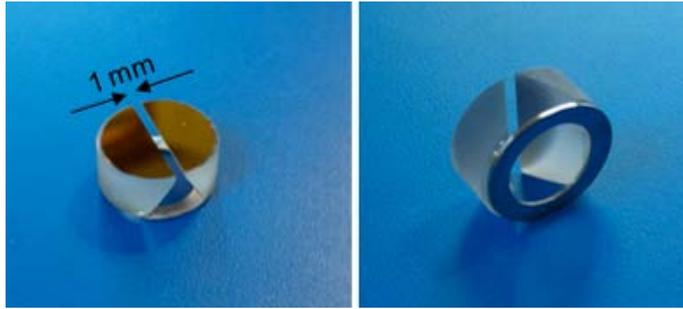

Figure 11 The two D mirrors are glued on an iron ring to take
them in position one with respect the other. Two reflecting planes are aligned on plane within 1'.

Changing the value of M-111 linear stage position it is possible to have the following OMs:

1) M-111 @ 0 mm (home position name: CC → 1+2 calib)
   The Light from sky is blocked by the back surface of the D shape mirrors. In this position the light from calibration input can be reflected by the bottom D shape mirror to the outgoing- fibers and the guider, along the input optical axis of the Preslit OIG box.

2) M-111 @ 1,7 mm (position name: CB → star fiber down + calib fiber up)
   The bottom fiber is feed by the light from sky while the top fiber is optically conjugated to the calibration beam. It is important to remember that the two parabolas M1 and M2 generate an inverting optical system (magnification from focus is 0.44): so the top fibers looks at the bottom f/11 focus.

3) M-111 @ 2,7 mm (position name: AC → star fiber up + calib fiber down)
   The top fiber is feed by the light from sky while the bottom fiber is optically conjugated to the calibration beam.

4) M-111 @ 15 mm (position name: AB → 1+2 sky)
   The D shape mirrors are placed in the bottom as far as possible from the mechanical axis of the input hole. The whole sky is visible by the guider: as already said the nominal FOV in this condition is 2X2 arcmin but it is partially vignetted by the mechanical mount that supports the D shape mirrors.

## 3. THE FIBERS-SPECTROMETER INTERFACE

The interface between the science fiber and Giano spectrometer is realized in the Preslit Giano Box, visible in Figure 12 and Figure 13. The box has the following dimensions:
606 X 456 X 140 mm and has a mass of 25 Kg. The whole system, composed by the box itself and the support structure, is attached to the Giano input flange by 12 M6 screws. The total mass is 35 Kg.

### 3.1 Opto-mechanical design

The 3D opto-mechanical design in shown in Figure 14 and the Zemax design in Figure 15.
The green beam is the optical path to feed the Image Slicer (SLC) actually used. The greater complexity of the operative optical beam is related only to the introduction of this last optical component (SLC).
The red beam is a backup optical path, if for some reasons we will decide to dismiss the Image slicer. So this backup beam is not deeply described in the present document

The starting point of the design of the Fiber-Spectrometer interface is the incoming minimum f/number that Giano Cryogenic optics can accept. This value is mechanically defined by the ratio of the distance cold stop - cold slit and the diameter of the cold stop itself: nominally it is f/9.5.
This choice comes from the hard definition of the fiber's f/number, due to the fiber focal ratio degradation (FRD, as used quality parameter for fiber ). The FRD varies with the incoming f/number and because of the polishing procedure of each single fiber.

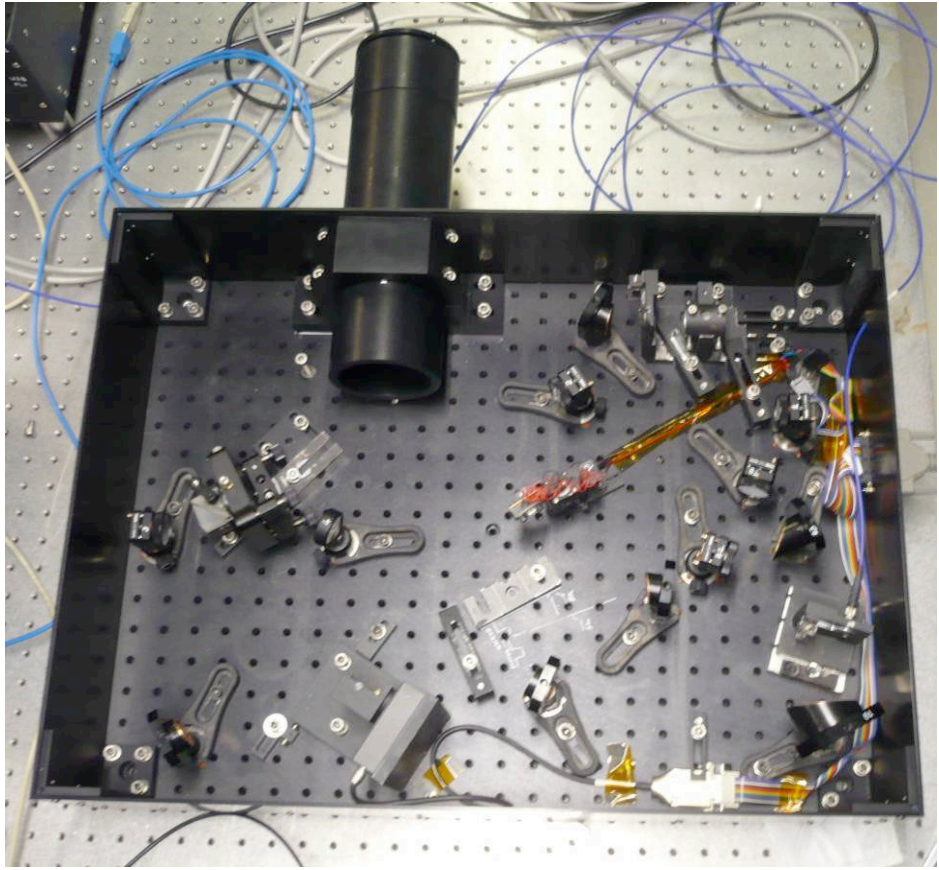
Figure 12 The inner of the Preslit Giano Box in laboratory before shipping to TNG.

The output beam coming from the Preslit OIG Box via the Science fiber optics, is collimated by the first Parabolic Mirror (P4), with 30° off axis angle and EFL=27,22 mm: the incoming beams from the two fibers are approximately f/4.9. The collimated beam is refocused by P5 (off axis Parabolic Mirror, off axis angle of 15° and EFL=654,99 mm). The generated focused beam has an f/# equal to f/120. The necessity to have a so big f/number is related to the difficulties to build an image slicer of small dimensions: with a so big f/number we have the images of the two fibers equal to 2,04 mm, being their real core diameter of 85 micron and the focal ratio of P5 and P4 of 24, and their center to center distance of 6,0 mm.

The beam is collected by P6 (15° off axis parabolic mirror, EFL=367,7 mm) that collimates the beam again. The last parabolic mirror, P7, generates the correct f/number in the correct position before entering in the matching lenses (L1-L2) placed near the input window of Giano Cryostat.
P7 has an off axis angle of 30° and an EFL=54,45 mm: it generates an f/16 beam, whose focus includes an image of the two sliced cores of the Science Fiber. Nominal diameter of the focus is 280 micron. This auxiliary focus is very useful to check quality beam before entering the Giano Cryostat.

After this f/16 focus the beam is folded by FM8 whose center is nominally placed on the axis of the cold stop of Giano. This mirror is placed on a rotator stage to permit the alignment of the focus with respect to the slit, that, we remember in Giano is placed quasi horizontal.

Lens L1 and L2 are the same matching lenses used the last year for the preliminary test of Giano and form a refocusing relay lens to well match the outcoming f/number of the preslit system with the nominal incoming f/number of the cold optics that if f/9.5: the nominal f/# is in fact f/10 and the diameter of the two foci on the slit is 175 micron, with respect to the slit dimension of 150 micron in the maximum resolving power position. The slit is mounted on a rotary stage that has several slits and other support alignment devices, but, actually, this wheel is fixed on the 150 micron slit and the motor driver is kept switched off permanently.

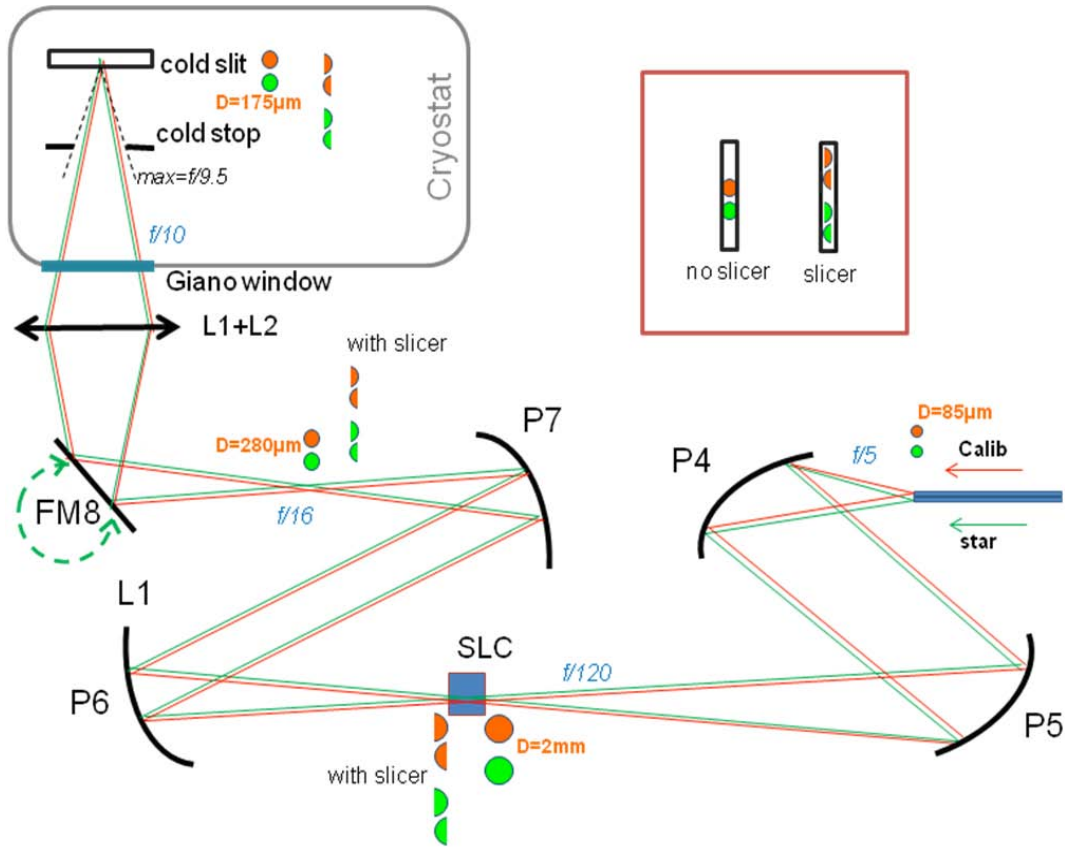

Figure 13 Schematic representation of the Fiber-Spectrometer interface, assembled into the Preslit Giano Box. P4-P7 are Parabolic mirrors, FM8 is the rotator folding mirror, SLC the sclicer. In the top right red box the matching between the slit and the foci is represented in case the slicer is used or not used.

In Figure 14 the 3D design of the Preslit Giano Box is shown: there they are shown the optical devices already mentioned. The device called "LED", that can be inserted into the optical collimated beam between P4 and P5, is the infrared LED already mentioned: it illuminates the fiber ends by P4 and it is used by FLI CCD to locate the fiber positions in the OIG box (Observation modality: fiber viewer).

### 3.2 Slit centering mechanism

The slit center mechanism is based on a commercial rotary stage (PI M116) that can control the vertical tilt angle of the folding mirror FM8 located before L1+L2 optical system. This plane mirror is nominally positioned vertically with respect to the optical bench and has its center located on the mechanical axis of the cold stop of Giano.

Because of the large field of view of L1+L2 reimaging system, the vertical position of the f/10 foci can be changed without affecting the optical quality of the foci themselves. The stage is based on a commercial Physik Instruments device (M-116) and has the appropriate angular resolution to realize a scan procedure of the slit with the purpose to maximize the signal on the Giano's detector.

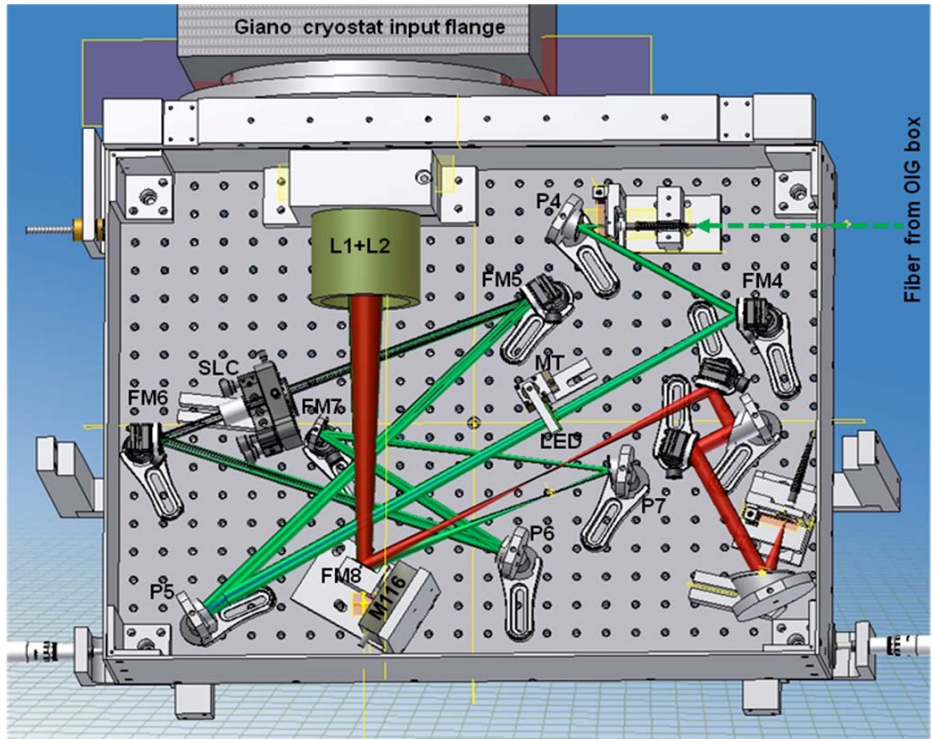

Figure 14 Inventor design of the Preslit Giano Box. In red the backup optical path, NOT USED and not described in the present document. In green the used one. P4-P7 are Parabolic Mirrors, FM4-FM8 are plane Folding Mirrors, L1-L2 custom Lenses, M116 a rotational stage, MT a motor to control the LED position and SLC is the Image SLiCer.

The whole mechanism is mounted on a mechanical support structure that can be manually rotated using two micrometer screws. The vertical rotary axis of this regulation is positioned on the center of FM8, within +/- 0.2 mm error.
This regulation is necessary to the horizontal fine positioning of the foci in the Giano cold slit, while, using the external box micrometers, you can roughly adjust the horizontal position of the foci by tilting and translating the whole Preslit Giano Box.

The translation rate of the foci with respect to the rotating angle of FM8 is 90 micron/arcmin being the dimension of the slit the following:

S075 = 1,120 x 0,140 mm ← Actually inserted
S050 = 1,120 x 0,093 mm [1]

That it means to cover the whole height of the S075 slit, it is necessary to rotate the M-116 of 1.6 arcmin. It can be done easily with great resolution using the PI stage:
- the minimum incremental motion of the stage is:    50 microrad (10,31 arcsec)
- unidirectional repeatability is:    12 microrad ( 2,50 arcsec)

The slit centering mechanism is one of the three devices that can move the image of the focus.
The second one is the slicer, that can be aligned using the six axes regulations of the commercial stage (by Thorlabs) on which it is mounted, and the third one is the output fiber connectors, that is a custom device.

---

[1] The Giano Detector pixel size is 18x18 micron. Each pixel on the input cold slit has a nominal dimension of 18 micron* 11/4.2 = 47 micron

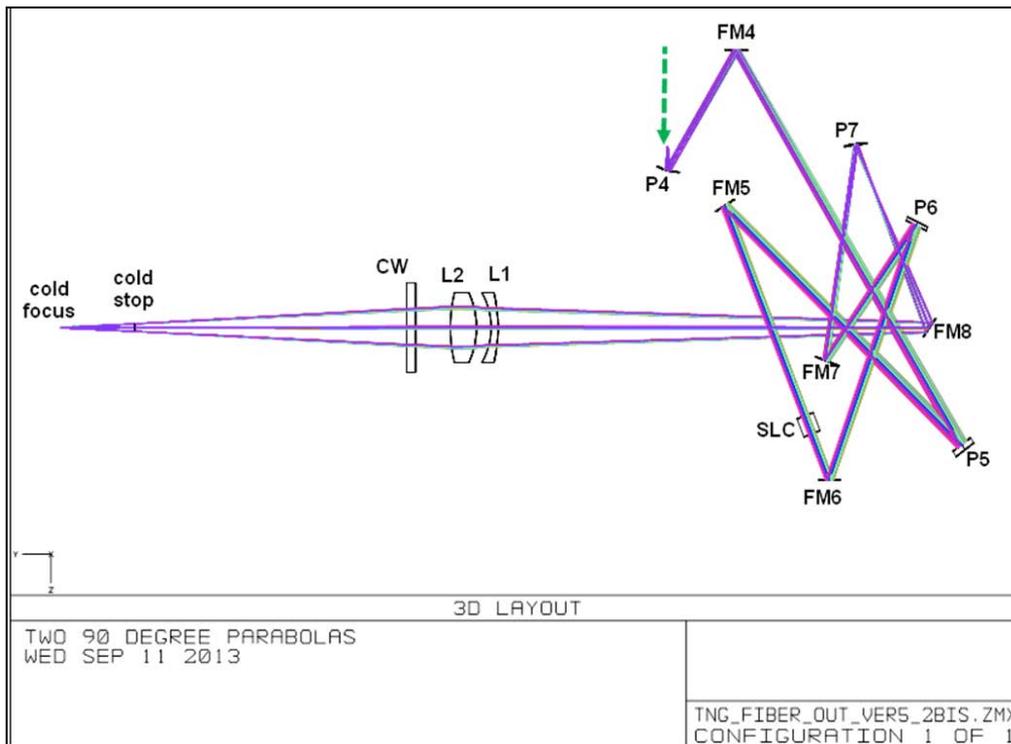
Figure 15 Zemax optical design for the Preslit Giano Box. P are commercial parabolas, FM folding mirrors, SLC is the image slicer

Using a "fine fiber angle regulation" it is possible to fine align the two foci on the slit in a range of +/- 3.5°: after this using the six axes regulation of the slicer it is possible to tune position and forms of the four half moons generated by the slicer.

### 3.3 The Image Slicer.

The image slicer is a particular optical component that is able to divide an input beam into two or more slices and to reassemble the single slice side by side in the output beam. In Figure 16 a photo of the Image Slicer is shown, where in red is represented the only input beam and in green the two sliced beams.

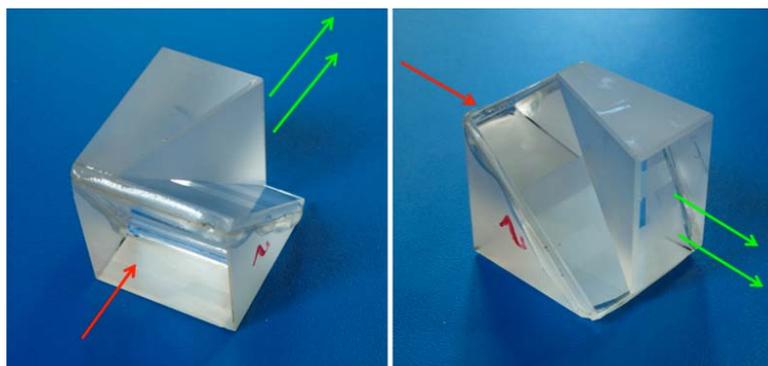
Figure 16 Photos of the Imge Slicer prisms. In red the input, in green the two outputs

The idea of the slicer is to recover the lost light that falls out of the slit as visible in Figure 17. The use of the slicer permits to rearrange the shape of the focus, without changing the Lagrange Invariance, recovering the lights that otherwise would be lost.

In case of Giano spectrometer there are two main slits, called S050 and S075, whose dimensions with respect to the dimension of the core of the science fiber (85 micron) are in one case too small and in the other a little bit too large. No other change was possible because of the pixel size dimension of the IR detector equal to 18x18 micron, that corresponds to 47x47 micron on input Cold Slit optical plane[2].

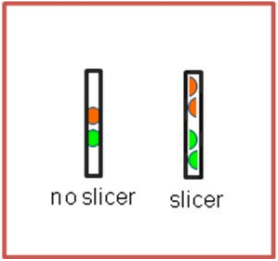

The core diameter of the two fibers is 85 micron, that are reimaged on the Cold Stop with a magnification of 2,06: so the diameter of the image is 175 micron.
Now we have to consider the dimension of the slit, already mentioned in the previous paragraph:

S075 = 1,120 x 0,140 mm (nominal collecting angle in sky of 0.75 arcsec) ← Actually inserted
S050 = 1,120 x 0,093 mm (nominal collecting angle in sky of 0.50 arcsec)

The possible alignments fiber-slit modes are four:
1. S050 slit without slicer: in this case we have a focus of 175 micron with respect to a slit of **93** micron: that means we lost approximately half light of the total transported by the fiber. In this case the spectrometer works "slit limited", as normally happens in standard spectrometers.

2. S075 slit without slicer: in this case the focus is again 175 micron and the slit 140 micron. We reduce the lost light, but so doing we are limited by the background noise in the K band, due to the bigger dimension of the slit. In fact the detector field of view permits to the detector to look at the two fibers image, but also to look at the zone around the two fibers, that is hot. This is the main source of photon noise: with respect to the use of S050 slit the background if 1,5 times more.

3. S050 slit with slicer: the focus diameter if 175 micron again, but because of the slicer we actually have four half moons with a height of 88 micron. The slit is 93 micron, so nominally it is possible to use this configuration, but it leads to a challenging alignment. A slightly rotation angle error in the positioning of fiber head, sends the half moons out of the slit, losing the image.

4. S075 slit with slicer: in this case we have not-round foci again, but two couples of half-moons (see Figure 17), one for each fiber. The height of the four semicircles is half a focus diameter that means 88 micron, with respect to the slit height of 140 micron. In this case we do not loose light in the matching, the alignment is simpler than case (3), but we have again a factor 1.5 of background noise more that using S050 slit. This is the configuration used for the Commissioning in July 2013.

It is interesting to point out that using the image slicer we don't need the slit from the point of view of the spectrograph resolution (but it continues to be important for the background light reduction): we work in a fiber limited condition, in which the image of the fiber is, itself, the slit and the spectral resolution is set by the slicer.

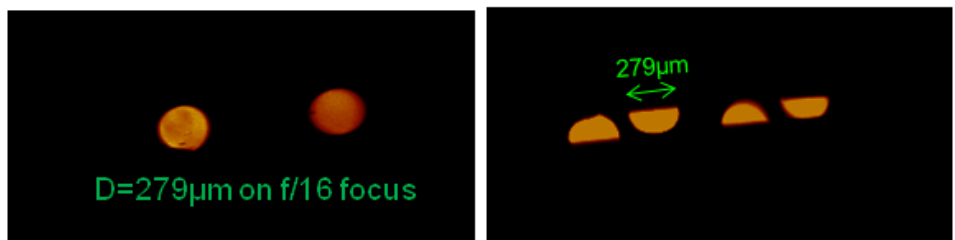

Figure 17 Images of the f/16 focus using calibration light, taken at TNG.
Left: without the use of the slicer. Right: with the slicer.

---

[2] The Giano Detector pixel size is 18x18 micron. Each pixel on the input cold slit has a nominal dimension of 18 micron* 11/4.2 = 47 micron

In Figure 18 a Zemax simulation of the Bowen Wallraven prism is shown. "Bowen Walraven image slicers use a thin glass plate where the light is transmitted along by total internal reflection. A base prism with a sharp edge is glued to the plate by molecular contact to cut the internal reflection of the transmitted beam. By choosing an appropriate configuration, the slices are arranged on a line simulating the slit of a spectrograph" [5].

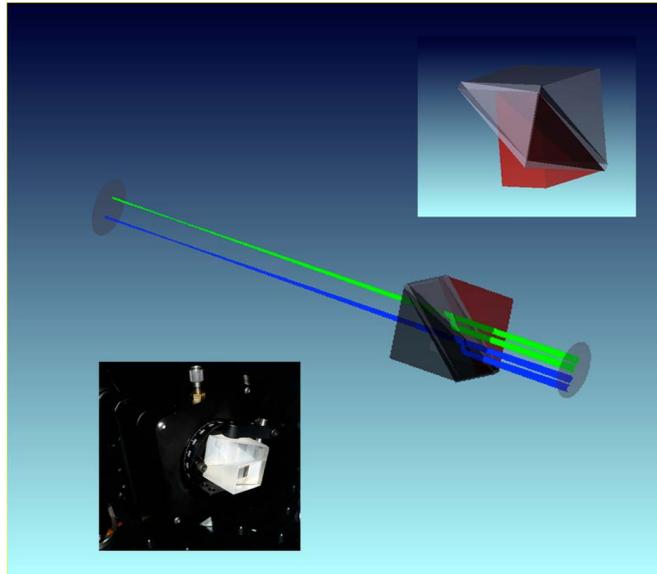

Figure 18 The Image Slicer of the Giano's preslit, based on a Bowen Wallraven prism.

Using the Fiber-Slit alignment modality number 4 we have measured an increasing signal in the GIANO's detector: the gain (G) of using the slicer is equal to 1.6 the case in which we don't use it (see Figure 19). On one hand the efficiency ($\eta$) of the image slicer optical element is 0.8, and this value can be found in an hypothetical GIANO working mode without the slit S050. On the other hand the efficiency of the slit S050 is 0.5, due to the required spectral resolution for GIANO and the consequent slit dimension with respect to the fiber image size. So this second value can be found in a GIANO working mode A without the image slicer and with the slit included.

The GIANO working case B, that includes both the image slicer and the slit S050, has a light efficiency equal to 0.8, because the presence of the image slicer ensures that no light loss happens through the slit way. The ratio between the light efficiency of the B working mode and A working mode gives the resulting gain coming from the image slicer use (G= 0.8/0.5=1.6).

### 3.4 Fiber scrambler mechanism

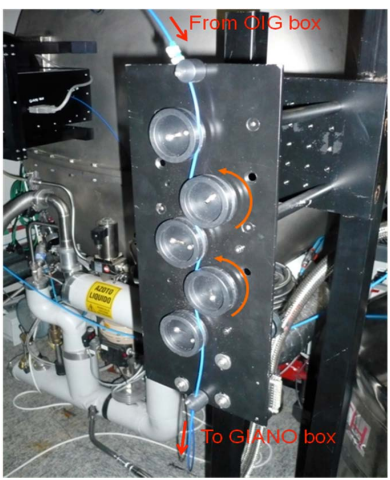

The fiber scrambler mechanism is able to mitigate the Modal Noise [6][7][8], that is usually present in an IR echellogram when a fiber optics is used to transport the telescope light to the spectrometer. It is based on a series of rotating disks that shakes the external fiber jacket. Because of the fragility of these ZBLAN fibers, the minimum static radius of torsion for these fibers is 50 mm: due to the dynamic way of use, we have doubled this specification.

The scrambler is positioned on the Nasmyth A platform near the Giano cryostat: the fibers from OIG box enters in the mechanism from the top side and exits to the bottom. The fiber has to pass in alternatively right and left of each rotating wheel, as visualize in the photo aside. During acquisition the nìmotor is automatically switched on by the control software.

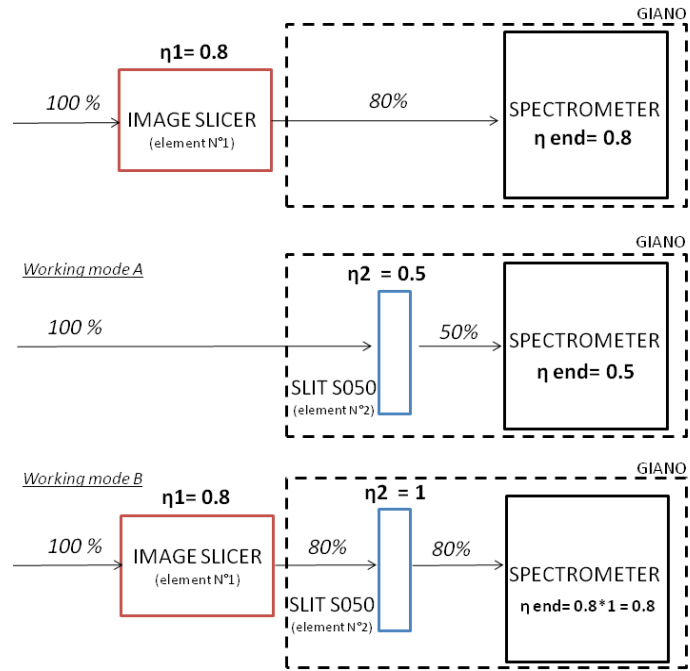

Figure 19 The efficiency of Giano is increased by the use of the Slice Image:
working in B modality the total efficiency is 0.8 with respect 0.5 of A modality.

## 4. MEASUREMENTS OF EFFICIENCY

This section summarizes the most significant measurements of absolute and relative efficiencies of GIANO and interfaces.

### 4.1 The absolute efficiency of the GIANO spectrometer

The measurements were performed using a commercial, certified black-body source (model CS500, manufactured by DIAS Infrared GmbH, emissivity=0.97). The measurement setup is shown in Figure 20.

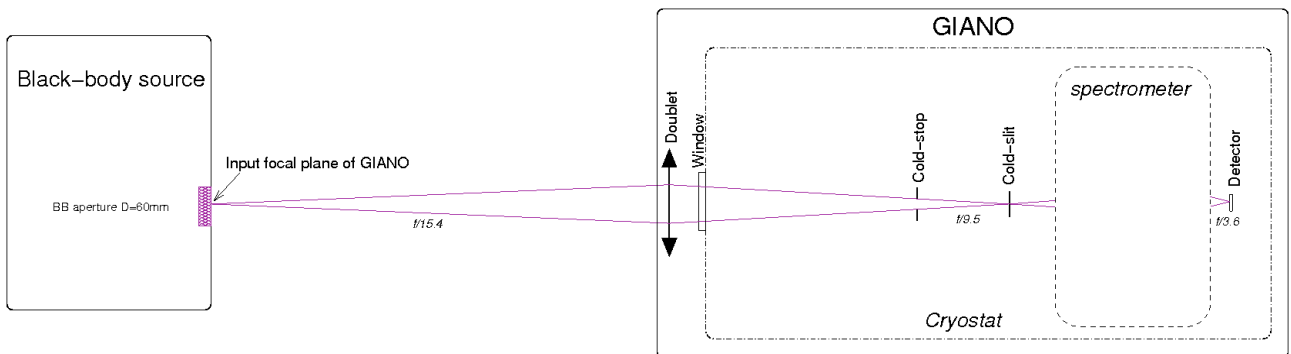

Figure 20 Schematic layout of the setup used to measure the absolute efficiency of the GIANO spectrometer.

The spectrometer was at operative conditions, with all the cryogenic parts stabilized at a temperature of about 80K. The only optics at ambient temperature were the lenses and window used to re-image the input focal plane of GIANO onto the cold-slit.

As cold-slit we used "S050" whose width and length (0.093 x 1.12 mm) projects onto 2 x 24 pixels of the detector. The spectral resolution of the data is determined by the slit width and amounts to $R=\lambda/\Delta\lambda=47,000$; thus each pixels sees a wavelength range $\Delta\lambda_{pix}=\lambda/47,000$ of the continuum emission from the calibration source.

The angular aperture of the light seen by the detector is defined by a cold-stop positioned before the cold-slit. The semi-angle of the cone is $\theta_{pix}=8.0^o$, equivalent to a focal ratio of F/3.6.

Measurements were performed with the black-body source stabilized at different temperatures, namely 108, 150, 200, 250, 300 and 350 $^o$C. For each temperature we used the data at a dozen of wavelengths, corresponding to the centers of the orders of the echellogram, i.e. the measurements correspond to the wavelengths of maximum efficiency of each order.

We included only the parts of the spectrum which were well detected and far from the saturation limit (<13,000 ADU/pix). The measured fluxes (ADU/pix) are summarized in Table 2; the values can be transformed in flux of photo-electrons using the conversion factor of GIANO array electronics: 2.2 el/ADU.

| Order | λ (μm) | Measured ADU/pix in 10s (multiply *0.22 to get photons/s/pix) | | | | | |
|---|---|---|---|---|---|---|---|
| | | T=108 $^o$C | T=150 $^o$C | T=200 $^o$C | T=250 $^o$C | T=300 $^o$C | T=350 $^o$C |
| 32 | 2.398 | 2090 | 9820 | - | - | - | - |
| 33 | 2.325 | 1480 | 7350 | - | - | - | - |
| 34 | 2.257 | 1015 | 5280 | - | - | - | - |
| 35 | 2.192 | 735 | 4050 | - | - | - | - |
| 36 | 2.131 | 520 | 3000 | - | - | - | - |
| 37 | 2.074 | 340 | 2100 | 11700 | - | - | - |
| 38 | 2.019 | 225 | 1440 | 8490 | - | - | - |
| 39 | 1.967 | 146 | 980 | 6070 | - | - | - |
| 40 | 1.918 | 92 | 665 | 4195 | - | - | - |
| 41 | 1.871 | - | 435 | 2950 | - | - | - |
| 42 | 1.827 | - | 295 | 2090 | 10050 | - | - |
| 43 | 1.784 | - | 202 | 1505 | 7490 | - | - |
| 44 | 1.743 | - | 135 | 1050 | 5490 | - | - |
| 45 | 1.705 | - | - | 725 | 3940 | - | - |
| 46 | 1.668 | - | - | 501 | 2895 | - | - |
| 47 | 1.632 | - | - | 340 | 1995 | 11550 | - |
| 48 | 1.598 | - | - | 230 | 1450 | 6330 | - |
| 49 | 1.565 | - | - | 155 | 1040 | 4705 | - |
| 50 | 1.534 | - | - | 110 | 745 | 3500 | 12450 |
| 51 | 1.504 | - | - | - | 520 | 2495 | 9595 |
| 52 | 1.475 | - | - | - | 380 | 1900 | 7205 |
| 53 | 1.447 | - | - | - | 270 | 1370 | 5395 |
| 54 | 1.420 | - | - | - | - | 1002 | 4250 |
| 55 | 1.394 | - | - | - | - | 740 | 3195 |
| 56 | 1.369 | - | - | - | - | - | 2400 |
| 57 | 1.345 | - | - | - | - | - | 1870 |
| 58 | 1.322 | - | - | - | - | - | 1410 |
| 59 | 1.299 | - | - | - | - | - | 1080 |
| 60 | 1.278 | - | - | - | - | - | 830 |
| 61 | 1.257 | - | - | - | - | - | 630 |

Table 2 Measured fluxes with GIANO spectrometer looking directly at a black-body source.

The photons-flux expected from the calibration source at a given temperature T is given by

$$\Phi_{\lambda,pix} = 2c\ \lambda^{-4} [exp(hc/\lambda kT)-1]^{-1}\ \varepsilon_{bb}\ A_{pix}\ \Delta\lambda_{pix}\ \Omega_{pix} \quad photons/s/pix$$

where

  $\varepsilon_{bb}$ = 0.97             emissivity of the calibration source
  $A_{pix}$ = 3.24 $10^{-6}$ cm$^2$         area of one pixel (pixel size =18μm)
  $\Omega_{pix}$ = $2\pi(1-cos\,\theta_{pix})$ = 0.061 sr    solid angle of light cone illuminating a pixel

The spectrometer absolute efficiency is the ratio between the measured flux of photo-electrons and $\Phi_{\lambda,pix}$. The results are summarized in Figure 21. The peak values of efficiency occur in the K-band and are about 22%, this value is very close to the 23% measured in Arcetri before shipping GIANO to the TNG. The decrease of efficiencies toward the shorter wavelengths is caused by the intrinsic drop of quantum-efficiency of the replaced detector which was sent to us in 2011, after the major failure of the original science-grade array.

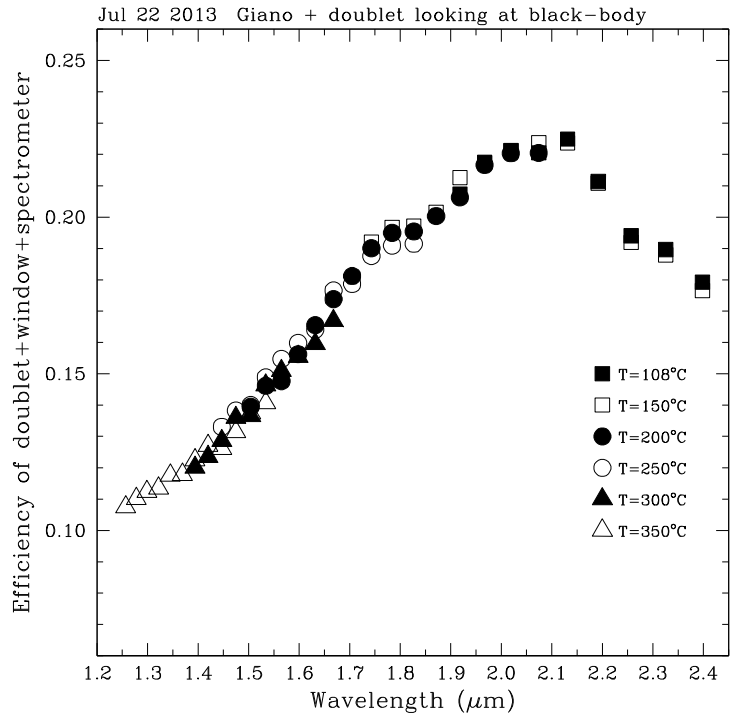

Figure 21 Absolute efficiencies of the GIANO spectrometer, without fibers-interfaces measured at TNG using a calibrated black-body source (see Figure 20)

The spectrometer was at operative conditions, with all the cryogenic parts stabilized at a temperature of about 80K. As cold-slit we used "S075" whose width and length (0.14 x 1.12 mm) projects onto 3 x 24 pixels of the detector.

## 4.2 The end-to-end absolute efficiency of GIANO

The measurements were performed at TNG in July 2013, using the same black-body source described in the previous section. The measurement setup is shown in Figure 22.

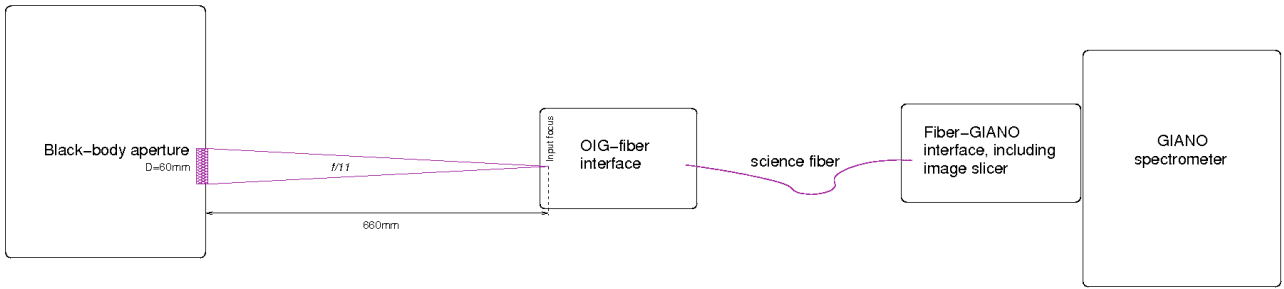

Figure 22 Schematic layout of the setup used to measure the end-to-end absolute efficiency of GIANO

This slit is larger than the image of the fibers. It is used as field-stop to minimize the thermal background at the longer wavelengths. The spectral resolution of the data is determined by the width of the image of the fiber produced by the image-slicer (see below).

The light was taken to the spectrometer through all the sub-systems which constitute the interface between GIANO and the telescope. In particular, the measurement included the same fiber and image-scrambler used for scientific observations.

The black-body source, whose aperture has a diameter of 60mm, was positioned at 660mm from the focus of the OIG-fiber interface. Thus the interface was illuminated with a beam aperture of F/11, identical to that of the telescope.

In this configuration, the spectrometer slit is the image of the fibers produced by the fiber interfaces. The image of two fibers, de-projected onto the entrance F/11 focus of the OIG-fiber interface, consists of 4 half-circles with radius 0.096 mm aligned along the direction perpendicular to the dispersion. Therefore, the total area illuminated by the black-body at F/11 is 0.058 mm$^2$.

The spectral resolution of the data is determined by the slit width of the fibers image. It amounts to R=$\lambda/\Delta\lambda$=47,000.

The measurements were performed with the calibration source stabilized at T=150 $^o$C. Data at other temperatures were not taken because of the long time necessary to change and stabilize the black-body temperature. The photo-electrons flux from the detector was determined as the flux per spectral resolution element (2 pixels), measured on the extracted 1D-spectrum which includes all the signals from the two fibers. The measured fluxes are summarized in Table 3; the values can be transformed in flux of photo-electrons using the conversion factor of GIANO array electronics: 2.2 el/ADU.

| Order | λ (μm) | Measured ADU per spectral resolution element in 10s |
|---|---|---|
|   |   | T=150 $^o$C |
| 32 | 2.398 | 66900 |
| 33 | 2.325 | 53200 |
| 34 | 2.257 | 39600 |
| 35 | 2.192 | 30400 |
| 36 | 2.131 | 23000 |
| 37 | 2.074 | 16400 |
| 38 | 2.019 | 11200 |
| 39 | 1.967 | 7510 |
| 40 | 1.918 | 4840 |
| 41 | 1.871 | 3220 |

Table 3 Measured fluxes with end-to-end GIANO system looking at a black-body source.

The photons-flux expected from the calibration source at a given temperature T is given by

$$\Phi_{\lambda,\text{fibe}} = 2c\,\lambda^{-4}[exp(hc/\lambda kT)-1]^{-1}\,\varepsilon_{bb}\,A_{fib}\,\Delta\lambda\,\Omega_{fib} \quad \text{photons/s}$$

where

$\varepsilon_{bb}$ = 0.97                                         emissivity of the calibration source

Δλ = λ/47,000                          width of spectral resolution element
$A_{fib}$ = 5.8 $10^{-4}$ cm$^2$               projected area of the fibers at F/11
$\Omega_{fib}$ = 2π[1−$cos$(0.5/11)] = 0.0065 sr     solid angle of F/11 light cone

The end-to-end absolute efficiency is the ratio between the measured flux of photo-electrons and $\Phi_{\lambda,fib}$. The results are summarized in Table 3 Measured fluxes with end-to-end GIANO system looking at a black-body source. The peak values of efficiency are about 0.091 which, once compared with the results shown in Figure 33 , indicates a total throughput of the fiber-interfaces of about 41%. This value is compatible with the measurement performed in Arcetri in June 2013, which yielded a total throughput of about 50% without the image-slicer.

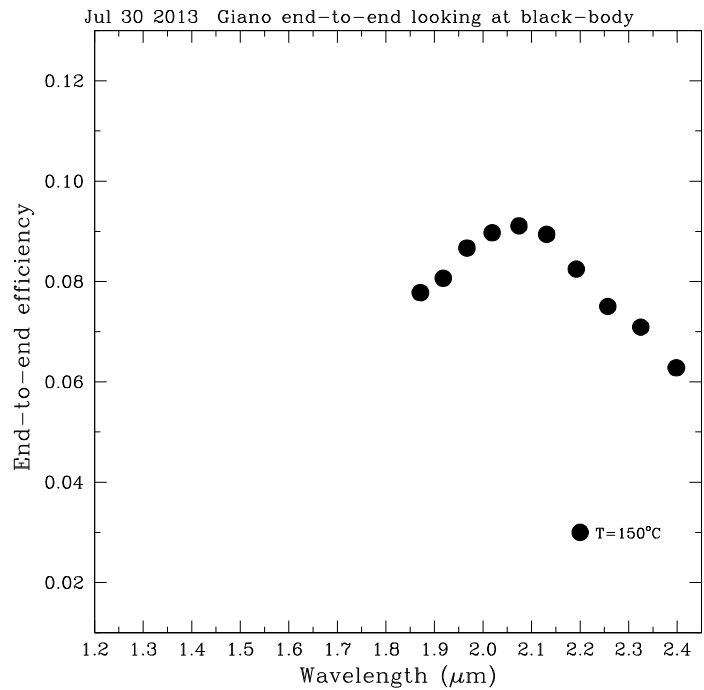

Figure 23 End-to-end absolute efficiencies of GIANO, including the ZBLAN fiber and the two prelist boxes.

## 4.3 The First Light of the Prelist System

The technical/scientific observations of stars with known fluxes, performed in the last run (October 2013) were used to compute the instrumental efficiencies. In the – relatively few – cases where the sky transparency and seeing conditions were acceptable, the derived throughput was a factor of about 2 below end-to-end value measured with a black body (Sect. 4.2). This is compatible with light losses at the entrance of the fiber caused by seeing or non-perfect centering of the object. Although a significant margin of improvement is still possible, e.g. by using fibers with better FRD and adding a proper fiber-viewer mechanism to better align the star with respect the input fiber core, we nonetheless conclude that the efficiency on sky is compatible with the measured value obtained with the black body and we conclude saying that the Giano instrument is fully ready for scientific operation.

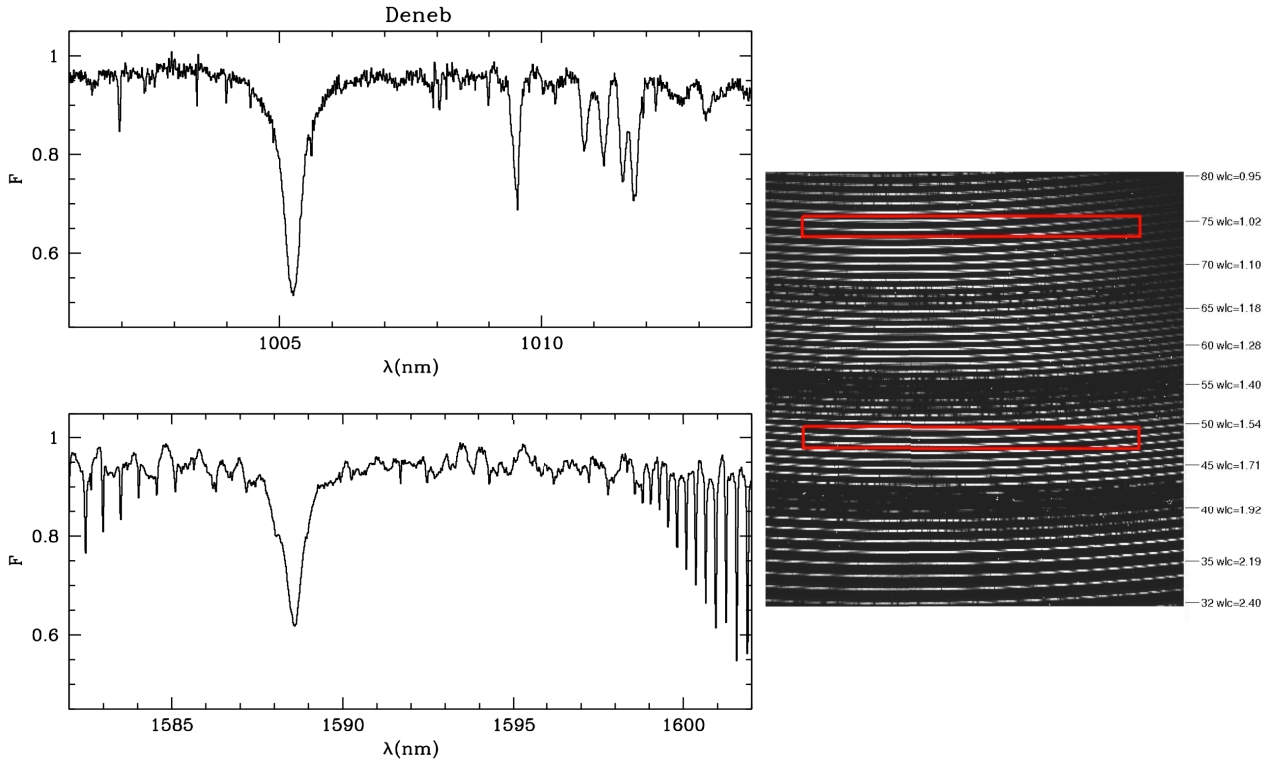

Figure 24 First light of GIANO Prelist system on 2013 at TNG: on the right the raw echellogram (200 s integration time), on the left the normalized spectra of the two red boxes are shown.

## 5. ACKNOWLEDGMENTS

This work was financially supported by INAF through the grants "TECNO-2011" and "TREX-2011".